\documentclass{revtex4}
\usepackage{graphicx}
\newcommand {\beq}{\begin{eqnarray}}
\newcommand {\eeq}{\end{eqnarray}}
\newcommand {\be}{\begin{equation}}
\newcommand {\ee}{\end{equation}}
\newcommand {\e}{\epsilon}
\def\bfm#1{\mbox{\boldmath $#1$}}
\newcommand{\BQ}{\begin{equation}}
\newcommand{\EQ}{\end{equation}}
\newcommand{\BQA}{\begin{eqnarray}}
\newcommand{\EQA}{\end{eqnarray}}

\newcommand{\kv}{\mbox{\boldmath $k$}}

\begin{document}
\begin{flushright}
NSF-ITP-02-45\\
hep-ph/0206043
\end{flushright}

\vspace{0.5cm}
\title{Color Superconductivity in Dense QCD\\
 and Structure of Cooper Pairs\footnote{Invited talk at the 
  Joint CSSM/JHF Workshop on Physics at Japan Hadron Facility 
  (March 14-21, Adelaide, 2002).}}

\author{Hiroaki Abuki$^{1)}$, Tetsuo Hatsuda$^{1),3)}$, 
        and Kazunori Itakura$^{2),3)}$\\{\hspace{1mm}}}

\affiliation{{ } \\$^{1)}$~Department of Physics, University of Tokyo, 
Tokyo 113-0033, Japan\\
$^{2)}$~RIKEN BNL Research Center, Brookhaven National Laboratory, 
              Upton, NY 11973, USA\\
$^{3)}$~Institute for Theoretical Physics, Universty of California, 
Santa Barbara, CA 93106-4030, USA
}


\begin{abstract}
The two-flavor color superconductivity is examined 
   over a   wide range of baryon density with a single model.  
  To study the structural change of Cooper pairs, quark correlation
   in the color superconductor is calculated both in the 
   momentum space and in the coordinate space.  
    At extremely high baryon density ($\sim O(10^{10} \rho_0))$,
   our model becomes equivalent to the usual perturbative QCD treatment
   and  the gap is shown to  have  a sharp peak near the Fermi surface  
   due to the weak-coupling nature of QCD.
   On the other hand,  the gap is a smooth function of the
   momentum  at lower densities ($\sim O(10 \rho_0)$)
  due to  strong color magnetic and electric interactions.
  The size of the Cooper pair is shown to become comparable
   to the averaged inter-quark distance at low densities,
    which indicates a crossover from BCS to BEC (Bose-Einstein
    condensation) of tightly bound Cooper pairs may take place 
   at low density.
 
\end{abstract}
 
\maketitle

\vspace{0.3cm}

\section{Introduction}

Because of the asymptotic freedom and the Debye screening in QCD,
  deconfined quark matter is expected to be realized 
  for baryon densities much larger than the normal nuclear matter
  density \cite{Collins_Perry}. 
This weak-coupling fermionic matter is, however, qualitatively
  different from a perturbed free Fermi
  gas system, if the temperature is low enough. 
This is because any attractive quark-quark 
 interaction in the cold quark matter  causes an instability of the Fermi 
 surface due to the formation of Cooper pairs and leads  to 
 the color superconducting phase
 \cite{BL_84,ARW,Review}.

 Current understanding on the color superconductivity has been
 based on two different theoretical approaches. One is an analysis
 by the Schwinger-Dyson  equation with perturbative one gluon 
 exchange, which is  valid in weak coupling limit at asymptotically 
 high densities \cite{Schaefer}.  
 The dominant contribution to the formation of Cooper pairs 
  comes from the collinear scattering through the 
 long range magnetic gluon exchange \cite{Son_98}.
 In such a weak-coupling regime,  formation of 
 Cooper pairs takes place only in a small region near the Fermi surface. 
 The other approach 
  is the mean-field approximation with the QCD inspired 4-Fermi 
  model  introduced to study lower density 
  regions\cite{ARW,Berges-Rajagopal}.
 In this approach, magnitude of the gap becomes as large as 100 MeV and 
  is almost constant in the vicinity of Fermi surface. 

 The main purpose of this talk is to discuss the superconducting
 gap over a wide range of baryon density with a single model and 
 to make a bridge between high and low density regimes \cite{AHI}.
 To this aim, we make an extensive analysis on the structural 
 change in spatial-momentum dependence of a Cooper pair in 2-flavor color
 superconductivity at zero temperature. 
 The momentum dependence of the gap, diffuseness of 
 the Fermi surface, quark-quark correlations in the superconductor
 are the characteristic quantities 
 reflecting the departure from the weak-coupling picture. 
 In particular, we show that the spatial size of Cooper pairs which is 
 calculated from the quark-quark correlations, indicates a clear deviation
 from the weak-coupling BCS theory.  
 More complete and detailed analysis was done in Ref.~\cite{AHI}.

\section{Gap equation with spatial momentum dependence}
 
Using the standard Nambu-Gor'kov formalism with a two component Dirac
spinor
  $\Psi=(\psi,\bar{\psi}^t)$, the quark self-energy $\Sigma(k_{\rho})$
 with the Minkowski 4-momentum $k_{\rho}$  satisfies 
\beq
  \Sigma(k_{\rho})&=&-i \int\frac{d^4q}{(2\pi)^4} \ g^2 \ 
 \Gamma_a^\mu S(q_{\rho})%
  \Gamma_b^\nu D_{\mu\nu}^{ab}(q_{\rho}-k_{\rho}),\label{DSeq}
\eeq
where $D^{ab}_{\mu\nu}$ is the gluon propagator in medium, 
$S(q_{\rho})$ is the full quark propagator, and
 $\Gamma^a_\mu$ is the quark-gluon vertex. 
We study the gap function in the flavor anti-symmetric, 
color anti-symmetric and $J=0^+$ channel (the most attractive 
channel within the one-gluon exchange model)\cite{Schaefer}:
\beq
\Delta(k_{\rho})=(\lambda_2 \tau_2 C\gamma_5)
 \left( \Delta_+(k_{\rho})\Lambda_+(\hat{\kv} ) +
    \Delta_-(k_{\rho})\Lambda_-(\hat{\kv} ) \right) ,
\eeq
where $\tau_2$ is the Pauli matrix acting on the flavor space, 
$\lambda_2$ is a color anti-symmetric Gell-Mann matrix,
and  $C$ is the charge conjugation. $\Lambda_\pm(\hat{\kv})\equiv%
(1\pm\hat{\kv}\cdot\bfm{\alpha})/2$ is the projector on 
 positive ($+$) and negative $(-)$ energy quarks.

For the  vertex in Eq.~(\ref{DSeq}), we use 
  $\Gamma^a_\mu=\mbox{diag.}(\gamma_\mu \lambda^a/2, %
 -(\gamma_\mu \lambda^a/2)^t)$.
For $g^2$ in Eq.~(\ref{DSeq}), we use a
 momentum dependent coupling $g^2(q,k)$ in the 
  ``improved ladder approximation'' \cite{Higashijima} 
{\bf (}$\beta_0=(11N_c-2N_f)/3=29/3${\bf )}:
\beq
  g^2(q,k)=\frac{16\pi^2}{\beta_0} \frac{1}{\ln \left(%
  (p^2_{\rm max} +p_c^2)/{\Lambda^2} \right)},
  \quad p_{\rm max}={\rm max}(q,k),
  \label{HM}
\eeq
where $p_c^2$ plays a role of a  phenomenological
 infrared regulator which prevents the coupling constant from being too 
  large at low momentum $q,k\sim \Lambda_{\rm QCD}$. At high momentum, 
  $g^2$ shows the same logarithmic behavior as the usual running coupling 
  with $\Lambda$ identified with $\Lambda_{\rm QCD}$.
The quark propagator in the improved ladder 
 approximation in the vacuum is known to have a high momentum behavior 
 consistent with that expected from the  renormalization group and the 
 operator product expansion. We adopt 
 $\Lambda$=400 MeV and $p_c^2$=1.5 $\Lambda^2$ which 
 are determined to reproduce the low energy meson properties for $N_f=2$ 
 vacuum~\cite{Ikeda}.

The gluon propagator   in Eq.~(\ref{DSeq}) 
in the Landau gauge reads,
\beq
  D_{\mu\nu}(k_{\rho}\, ;\, k_0 < |\bfm{k}|)=
-\frac{P^{\rm T}_{\mu\nu}}{\bfm{k}^2+iM^2 |k_0|/|\bfm{k}|}%
   -\frac{P^{\rm L}_{\mu\nu}}{\bfm{k}^2+m_{\rm D}^2}, \label{propagator}
\eeq
where $P^{\rm T,L}_{\mu\nu}$ are the transverse and longitudinal
projectors. 
 The longitudinal part of the propagator (the electric part) 
 has a static screening by the Debye mass
 $m_{\rm D}^2 = (N_f/2\pi^2) g^2 \mu^2 $, while the
 transverse part (the magnetic part) has dynamical screening by the 
 Landau damping $M^2=(\pi/4)m_{\rm D}^2$.
 This form is a quasi-static
 approximation of the full gluon propagator in the sense
 that only the leading frequency dependence is considered~\cite{Iida}.

\subsection{Full gap equation}

The gap equation under the approximations shown above
 is obtained from the 2-1 element of the Schwinger-Dyson 
 (matrix) equation (\ref{DSeq}) which reads 
\beq
\Delta_\pm(k) =
         \int_0^\infty dq\  V_\pm(q,k;\e_q^+,\e_k^{\pm})
           \frac{\Delta_+(q)}{\sqrt{E_+(q)^2+|\Delta_+(q) |^2 }} \nonumber \\
       \ \ \ \ \ \ \ \ \ \ \ \ \ \ \ \
  +\int_0^\infty dq\  V_\mp(q,k;\e_q^-,\e_k^{\pm})
           \frac{\Delta_-(q)}{\sqrt{E_-(q)^2+|\Delta_-(q) |^2 }}.
      \label{full-gap}
\eeq
Here we used a simplified notation: 
$\Delta_{\pm}(\e_k^{\pm},k) \rightarrow \Delta_{\pm}(k)$ with
$E_{\pm}(q) \equiv q \mp \mu$ and $\e_k^{\pm}$
 being  the quasi-particle energy
 as a solution of $(\e_q^{\pm})^2 = 
        E_{\pm}^2(q) + \Delta_{\pm}^2(\e_q^{\pm},q)$.
 The explicit form of $V_\pm$ is given in Ref.~\cite{AHI}.

At extremely high density, the Cooper pairing is expected to take
  place only near the Fermi surface due to weak coupling property.
 In this case,  we can safely neglect the antiquark-pole contribution for
 calculating $\Delta_+$.
 Furthermore, one may replace the momentum dependent coupling constant
  (\ref{HM}) by that on the Fermi surface $g^2(\mu,\mu)$.
On the other hand, at low densities,
  sizable diffusion of the Fermi surface occurs and the
  weak-coupling  approximation   is not justified, and we need to solve
  the coupled gap equations  Eq.~(\ref{full-gap}) numerically.

\subsection{Occupation number, correlation function and coherence length}

To clarify the structural change of the color superconductor
 from high to low densities,
 it is useful to examine the following  physical quantities.


\noindent
{\bf [I]} {\it The quark and antiquark occupation number}
    which
    is related to the diagonal (1-1) element of the 
 quark propagator, 
$\langle\psi^{\dagger b}_{j}(t,\bfm{y})\psi^a_i(t,\bfm{x})
\rangle_{\rm super}$, in the   Nambu-Gor'kov formalism;
\be
n_{\pm}^{1,2}(q)
=\frac{1}{2}\left(1-\frac{E_{\pm}(q)}{\sqrt{E_{\pm}(q)^2
 +|\Delta_{\pm}(q)|^2}}\right),\  n_+^3(q)=\theta(\mu- q), \ 
 n_-^3(q)=0,\, 
 \label{occupation}
\ee
where the superscripts (1, 2 and 3) stand for color indices.
Since the third axis in the color space
  is chosen to break the color symmetry,  quarks
 with  the third color remains ungapped.


\noindent
{\bf [II]} {\it  The
 $q$-$q$ and $\bar{q}$-$\bar{q}$ correlation functions in  
   momentum space $\hat{\varphi}_{\pm}(q)$
  and in coordinate
 space ${\varphi}_{\pm}(r)$}:
 They reflect the internal structure of the  Cooper pairs
 in color superconductor. 
  These correlations  are defined through the off-diagonal (1-2) element
  of the quark propagator, 
  $\langle\psi^a_i(t,\bfm{x})\psi^b_j(t,\bfm{y})\rangle_{\rm super}$;
\beq 
\hat{\varphi}_{\pm}(q)=  
\frac{\Delta_\pm(q)}{2\sqrt{E_\pm (q)^2
+|\Delta_\pm (q)|^2}}, \quad
\varphi_{\pm}(r)= N \int\frac{d^3\bfm{q}}{(2\pi)^3}
\  \hat{\varphi}_{\pm}(q)\  e^{i  \bfm{q}\bfm{r}}, \label{correlation}
\eeq
\vspace{0.2cm}
where $N$ is a normalization constant determined by
$\int d^3r  |\varphi_+ (r) |^2 =1$.


\noindent
{\bf [III]} {\it The coherence length
 $\xi_{\rm c}$  characterizing the 
 typical size of a Cooper pair}:
  It is defined simply as a root mean square radius
 of $\varphi_{+}(r)$: 
\beq
  \xi_{\rm c}^2 
 = \frac{\int d^3r \  r^2 |\varphi_+ (r) |^2}{ \int d^3r 
   |\varphi_+ (r) |^2}
 =
\frac{\int_0^\infty dk\ k^2
             \left| d\hat{\varphi}_{+}(k)/ dk  \right|^2}{
 \int_0^\infty dk\ k^2\left|\hat{\varphi}_{+}(k)
    \right|^2 } .
  \label{coherence-length}
\eeq
It can be shown\cite{AHI} that the quark correlation
  $\varphi_+(r)$ in the weak-coupling limit behaves as 
\beq
\varphi_+(r \rightarrow \infty )\propto  
 \frac{ \sin (\mu r)}{(\mu r) ^{3/2}} \cdot
  e^{-r/(\pi \xi_{\rm p})},\label{asymptotic_WF}
\eeq
 where the Pippard length is given by $\xi_{\rm p}=(\pi\Delta_+(\mu))^{-1}$. 

In a typical type-I superconductor  in metals, the Pippard length is of
  the semi-macroscopic order $\xi_{\rm p} \sim 10^{-4}$ cm, whereas
  inverse
  Fermi momentum is of the microscopic order $k_{\rm F}^{-1}\sim 10^{-8}$
  cm.
 Besides, there is another scale $\omega_{\rm D}$, the Debye cutoff, 
  which limits the range of attractive interactions to be just around the 
  Fermi surface $|k-k_{\rm F}|<\omega_{\rm D}$.
 The inverse of the Debye cutoff is in between the two scales above:  
   $\omega_{\rm D}^{-1} \sim 10^{-6}$ cm.
 Therefore there is a clear scale  hierarchy,
   $\Delta \ll \omega_{\rm D} \ll k_{\rm F}$.
 On the other hand, since there is no intrinsic  scale $\omega_{\rm D}$ 
  in QCD, scale hierarchy at extremely high density simply reads 
    $\Delta\sim \mu e^{-c/g} \ll k_{\rm F}\sim \mu$.  
 At lower densities, however,  such scale separation becomes
    questionable for $g$ is not small.


\section{Momentum dependent gap from high to low densities}

\begin{figure}[htbp]
\vspace{2mm}
  \centerline{
 \begin{minipage}{0.49\textwidth}
 \includegraphics[scale=0.44]{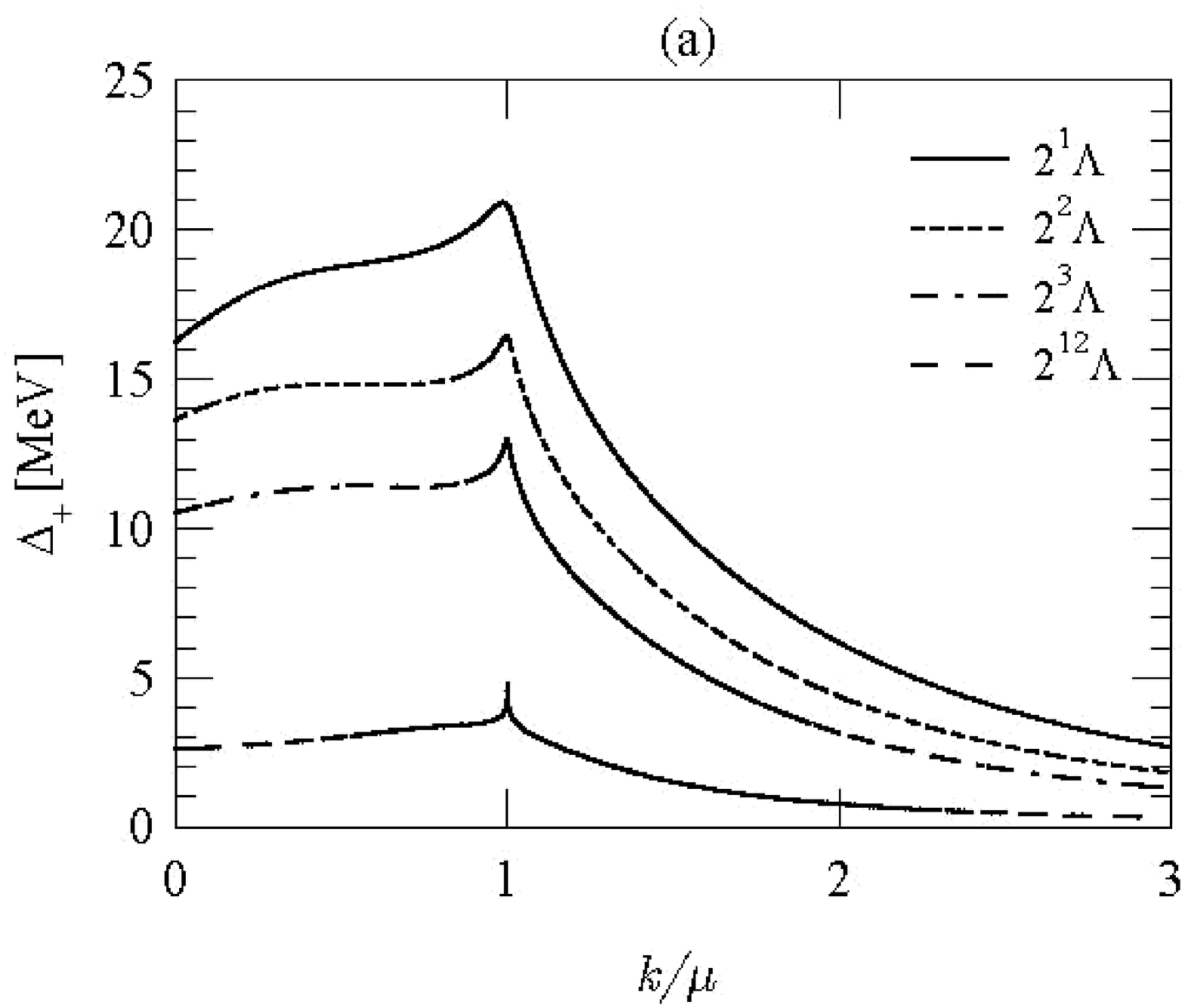} 
 \end{minipage}%
 \hfill%
 \begin{minipage}{0.49\textwidth}
  \includegraphics[scale=0.59]{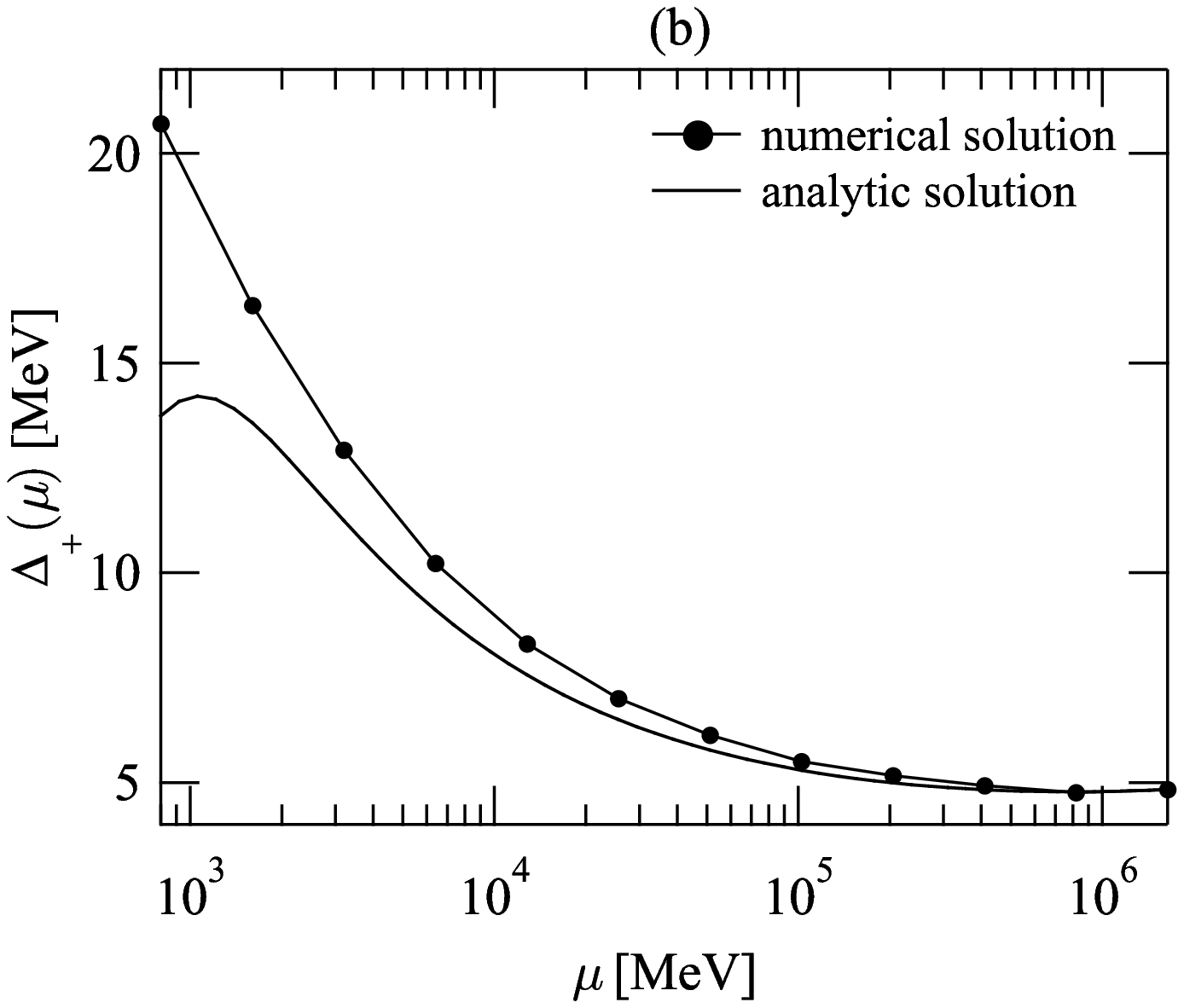} 
 \end{minipage}%
  }
  \caption{%
    (a) $\Delta_+(k)$ as a function of $k/\mu$ for various densities 
        $\mu =2^n\Lambda $ with $n=1,2,3,12$. ($\Lambda=$400MeV).
        All the calculations are done with the momentum dependent
        vertex $g^2(q,k)$ and with the antiquark-pole contribution.
    (b) Chemical potential dependence of the gap $\Delta_+(k=\mu)$  
        in the full calculation compared with the analytic result 
        which is normalized at the highest density $\mu = 2^{12} \Lambda$.}
  \label{fig:6}
\end{figure}

In Fig.~\ref{fig:6}(a), we show the gap $\Delta_+(k)$ as a solution of
   the full gap equation  for a wide range of densities.
   Since the actual position of the Fermi surface moves as we vary 
   the density, we use  $k/\mu$ as a horizontal axis in the figure 
   which helps us to understand the change of global behavior.
 The figure shows that the sharp peak at high density gradually gets
   broadened and simultaneously 
    the magnitude of the gap increases as we decrease the density.

  The characteristic features at {\it high density} are 
  (i) there is a sharp peak at the Fermi surface,
      and 
  (ii) the gap decays rapidly but 
  is nonzero for momentum far away from the Fermi surface.
  The property (i) is similar to the standard BCS superconductivity
 but (ii) is not, due to the absence of intrinsic
  ultraviolet  Debye-cutoff of the gluonic interaction in QCD.
  As for the magnitude of the gap at high density,
  if one estimates the several contributions to the kernel $V_\pm$ in 
  the gap equation (\ref{full-gap}) separately, 
  one finds that the color-electric interaction enhances the gap 
  considerably. 
 This may be understood as follows:
  In the coordinate space, the Debye-screened electric interaction is 
  a Yukawa potential. Such a short-range interaction alone can form
   only a loosely bound Cooper pairs and a very small gap.  However, 
    if the magnetic and electric interactions coexist,  
   small size Cooper pairs are formed primarily by the long-range magnetic
   interaction. Then,  even the short-range electric interaction becomes
   effective to  generate further attraction between the quarks.

The characteristic features at {\it low density} are that 
 (iii) the sharp peak at the Fermi surface disappears, and
 (iv)  all the contributions neglected
 in the weak-coupling limit are actually not negligible.
 A close look at the effects of each contributions to the gap leads us 
 to a conclusion that the color superconductivity
 at low density is not a phenomenon just around the Fermi surface.
 This is confirmed by computing the occupation number following 
 eq.~(\ref{occupation}). The result shows that the Fermi surface is diffuse 
 substantially at low density.

 In Fig.~\ref{fig:6}(b), the gap at the Fermi surface $\Delta_+(\mu)$ 
  is shown as a function of the chemical potential.  
  It decreases monotonically as $\mu$ 
  increases as is shown on the figure, 
  but turns into an increase at much higher density $\mu > 10^6 $ MeV.
  An analytic solution based on the weak coupling approximation
  (includes only the magnetic gluons, fixed coupling, and ignores 
   the anti-quark pole contribution) is also shown 
 in Fig.~\ref{fig:6}(b). The magnitude of the analytic solution is 
 normalized to the numerical solution at the highest 
  density $\mu=2^{12}\Lambda \simeq 1.6 \times 10^6 {\rm MeV}$.
 At high density, 
  $\mu$-dependence of  the numerical result is in good agreement with the 
 analytic form which has  a parametric dependence 
  $\Delta_+(\mu) \propto  g^{-5}\mu\exp(-3\pi^2/\sqrt{2}g)$ with  
$g^2= g^2(\mu,\mu)$.
 On the other hand, the difference of the two curves at low density implies
 the 
  failure of the weak-coupling approximation.

\section{Correlation function and coherence length}

  One of the advantages  of treating the momentum dependent gap is that 
    we are able to calculate the correlation function which physically
    corresponds to the ``wavefunction'' of the Cooper pair. Such
    correlations  have been first
     studied in Ref.~\cite{Matsuzaki} in the 
     context of the color superconductivity (but with much simpler model 
     than eq.~(\ref{full-gap}), and only at lower density).
 
\begin{figure}[htbp]
 \vspace{2mm}
 \begin{minipage}{0.49\textwidth}
 \includegraphics[scale=0.6]{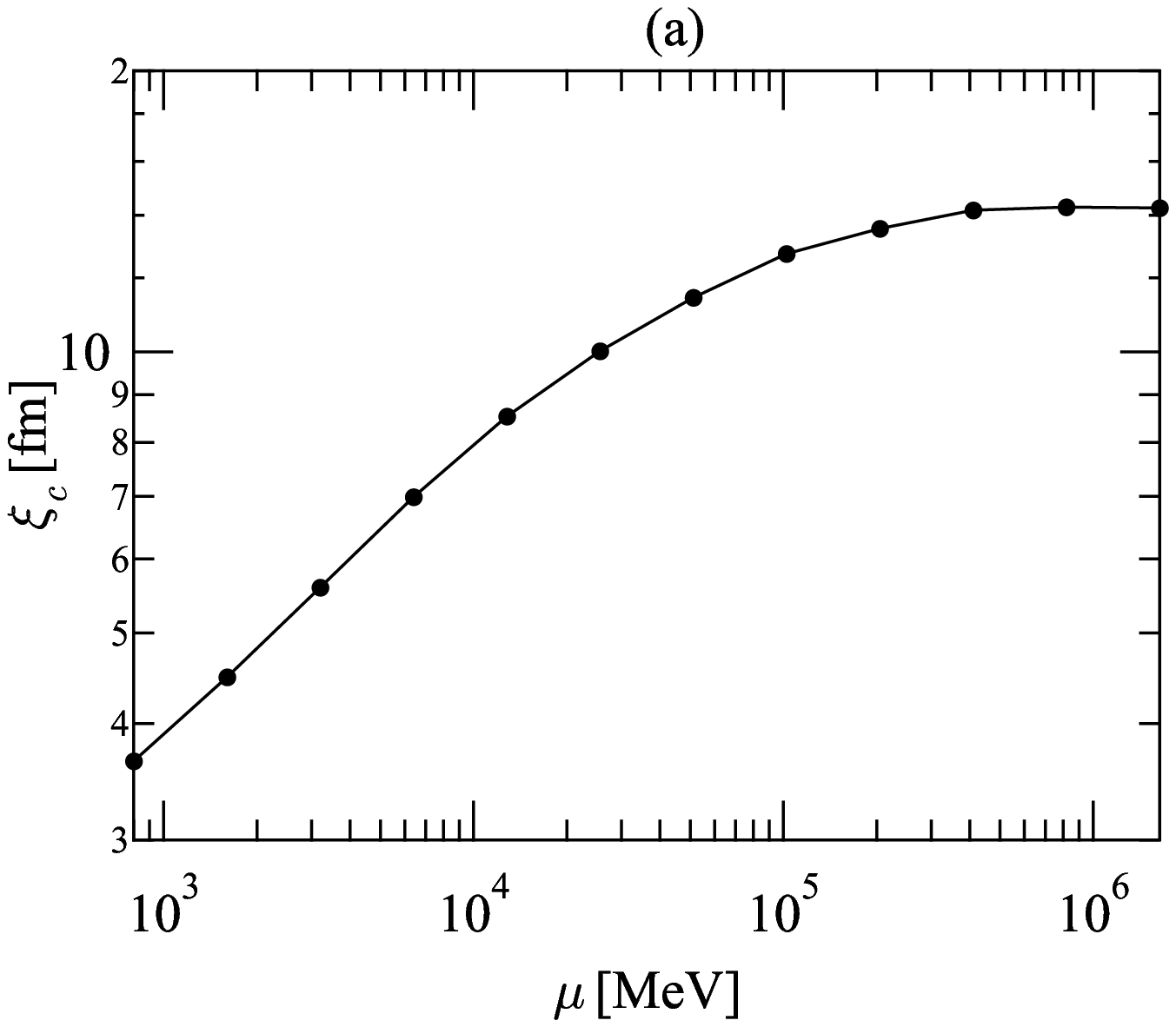} 
 \end{minipage}%
 \hfill%
 \begin{minipage}{0.49\textwidth}
  \includegraphics[scale=0.6]{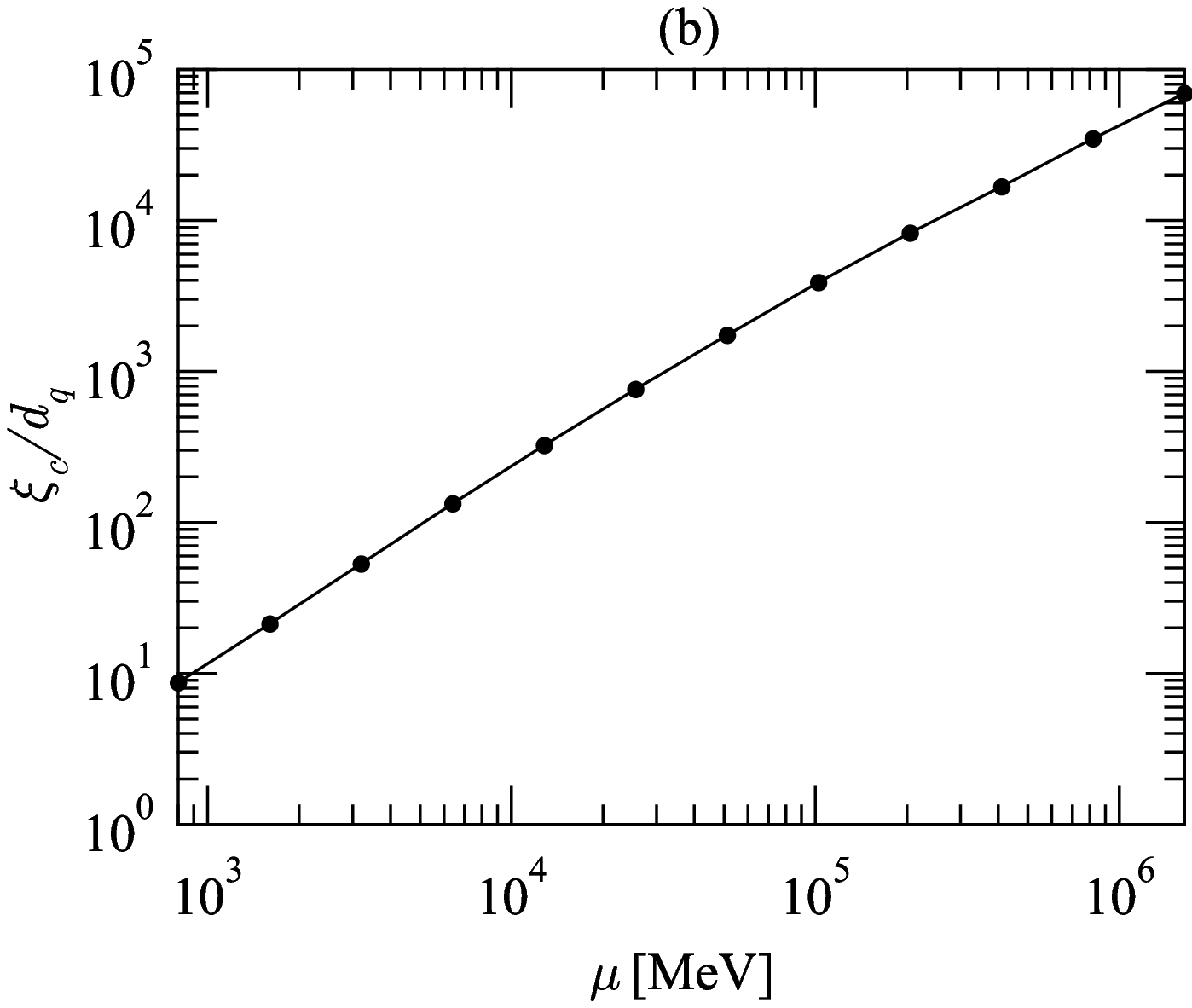} 
 \end{minipage}%
  \caption{(a): Density dependence of the coherence length.
    (b): Ratio of the coherence length $\xi_{\rm c}$  and the
          average inter quark distance $d_q$ as a function of the
           chemical potential.}
  \label{Coh}
\end{figure}

 Figure~\ref{Coh}(a) shows the coherence length $\xi_c$ of a quark Cooper 
    pair defined as the root mean square radius of the correlation function
    [see Eq.~(\ref{coherence-length})].
 The size of a Cooper pair becomes smaller as we go to lower     densities. 
 This  tendency   is understood by the 
 behavior of the  Pippard length $\xi_{\rm p}=1/\pi\Delta_+(\mu)$ 
  (which gives a rough estimate of the coherence length)
   together with the behavior of $\Delta_+(\mu)$ shown in 
 Fig.~\ref{fig:6}(b).

 Also,  the Cooper pair becomes 
   smaller  as density increases beyond  $\mu=2^{12}\Lambda$.
  However, it does not necessarily imply the existence of tightly bound 
    Cooper pairs. 
 In fact,
 the size of a Cooper pair  makes sense only   in comparison to the
  typical length scale of the system, namely
 the  averaged inter-quark distance $d_q$ for free quarks defined  as
$$
  d_q=\left(\frac{\pi^2}{2}\right)^{1/3}\frac{1}{\mu}.
$$
As we go to higher densities,   the ratio $\xi_{\rm c}/d_q$ 
 increases monotonically as shown   in 
    Fig.~\ref{Coh}(b). Namely  loosely bound Cooper pairs similar
      to the BCS superconductivity
    in metals are formed at extremely high densities.
(Recall that the typical ratio in superconductivity is $k_{\rm F}/\Delta 
\sim 10^{4}$)

 At the lowest density in Fig.~\ref{Coh}, 
 the size of the Cooper pair is less than 4 fm and 
  the  ratio $\xi_{\rm c}/d_q$ is less than $10$.
 This is not similar to the usual BCS system.
 The transition from  $\xi_{\rm c}/d_q \gg 1 $ to $\xi_{\rm c}/d_q \sim 1$
    as $\mu$ decreases is analogous to the crossover from the BCS-type
     superconductor to the Bose-Einstein condensation (BEC) of 
    tightly bound Cooper pairs 
    \cite{BEC}.
    Our result here suggests that the quark matter possibly realized in
    the core of neutron stars may be rather like the BEC of tightly bound
    Cooper pairs.

\begin{figure}[htb]
  \centerline{
 \includegraphics[scale=0.45]{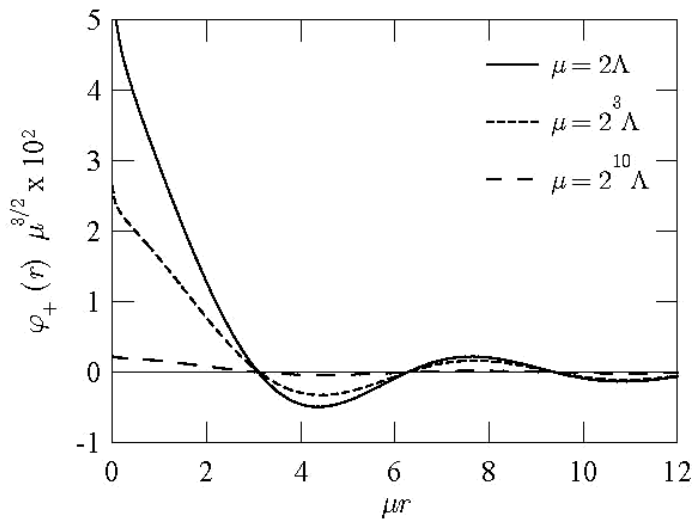} 
  }
  \caption{The quark-quark  correlation function $\varphi_+(r)$ 
  in the coordinate space  $\varphi_+(r)$  for  
  several different  chemical potentials. $\varphi_+(r)$ is normalized to
   be unity.
 }
 \label{Spatial_WF}
\end{figure}

For better understanding of the internal structure of the quark Cooper pair,
  let us consider the correlation function in the coordinate space.
 Fig.~\ref{Spatial_WF} shows the spatial correlation 
    of a Cooper pair at various chemical potentials 
  normalized as
$\int d^3{r}|\varphi_+(r)|^2=1.$
 
 At high density, most of the quarks participating in forming  
   a Cooper pair have the Fermi momentum  
   $k_{\rm F}=\mu$ giving a sharp peak in the momentum space correlation.
 In the coordinate space, this corresponds to an oscillatory
   distribution   with 
  a wavelength $\lambda=1/\mu$ without much structure
   near the origin. (The oscillation  is also evident from the
   factor $\sin (\mu r)$ in the 
  approximate correlation function Eq.~(\ref{asymptotic_WF}).) 
   At lower densities, accumulation of the 
    correlation near the origin in the coordinate space
     is much more prominent in Fig.~\ref{Spatial_WF}. This  
 implies a localized Cooper pair composed of quarks with various
  momentum.


\section{Summary and Discussion}

 We have studied the spatial-momentum dependence of a 
    superconducting gap and the structure of the Cooper pairs in 
    two-flavor color superconductivity, using a single model for a very 
    wide region of density.
 Nontrivial momentum dependence of the gap manifests itself at low 
    densities, where relatively large QCD coupling  allows the Cooper
   pairing to take place in a wide region around the Fermi surface.
 Note that our results can be easily extended to $N_f=3$ case.
 Our results imply that the quark matter which might exist in the core
    of neutron stars or in the quark stars could be  rather different 
    from that expected from the weak-coupling BCS picture
    and could be more like a BEC of tightly bound
    Cooper pairs.  Study of the finite $T$ phase transition of this 
     strong coupling system is currently under investigation.

\section*{Acknowledgments}
T.H. would like to thank A. W. Thomas, A. G. Williams, D. Leinweber
 and the members of the local organizing committee 
   for giving him an opportunity to give a talk at the 
  Joint CSSM/JHF Workshop on Physics at Japan Hadron Facility 
  (March 14-21, Adelaide, 2002). This research was supported
 in part by the National Science Foundation under Grant No. 
 PHY99-07949.

\end{document}